\documentclass[twocolumn,pra,aps,showpacs,nofootinbib,superscriptaddress]{revtex4}
\usepackage{xspace,amsmath,amsfonts,amsthm,amssymb,amsbsy,graphicx,color,setspace}
\pdfoutput=1
\newcommand{\name}{\mbox}
\newcommand{\ket}[1]{| #1\rangle}
\newcommand{\bra}[1]{\langle #1 |}
\newcommand{\braket}[2]{\langle #1 | #2 \rangle}
\newcommand{\ketbra}[2]{\ket{#1}\!\bra{#2}}
\newcommand{\pbfrac}[2]{\mbox{$\mbox{}^{#1}\!/_{#2}$}}
\newcommand{\guess}{b}
\newcommand{\basis}{{\mathcal B}}
\newcommand{\eps}{\varepsilon}
\newcommand{\pr}{\mbox{Prob}}
\newcommand{\Tr}{\mbox{Tr}}
\newcommand{\D}{\mbox{D}}

\allowdisplaybreaks[3]
\begin{document}

\title{Fair Loss-Tolerant Quantum Coin Flipping}
\date{\today}
\author{Guido Berl\'\i n}\email{berlingu@iro.umontreal.ca}
\affiliation{D\'epartement d'informatique et de recherche op\'erationnelle,
Universit\'e de Montr\'eal \\ 
C.P.~6128, Succursale Centre-Ville, Montr\'eal, Qu\'ebec, H3C~3J7~Canada}
\author{Gilles Brassard}\email{brassard@iro.umontreal.ca}
\affiliation{D\'epartement d'informatique et de recherche op\'erationnelle,
Universit\'e de Montr\'eal \\ 
C.P.~6128, Succursale Centre-Ville, Montr\'eal, Qu\'ebec, H3C~3J7~Canada}
\author{F\'elix Bussi\`eres}\email{felix.bussieres@polymtl.ca}
\affiliation{Laboratoire des fibres optiques,
D\'epartement de g\'enie physique, \'Ecole Polytechnique de Montr\'eal \\ 
C.P.~6079, Succursale Centre-Ville,
Montr\'eal, Qu\'ebec, H3C~3A7~Canada}
\affiliation{Institute for Quantum Information Science and Department of Physics and Astronomy \\ 
University of Calgary, 2500 University Drive NW, Calgary, Alberta, T2N~1N4~Canada}
\author{Nicolas Godbout}\email{nicolas.godbout@polymtl.ca}
\affiliation{Laboratoire des fibres optiques,
D\'epartement de g\'enie physique, \'Ecole Polytechnique de Montr\'eal \\ 
C.P.~6079, Succursale Centre-Ville,
Montr\'eal, Qu\'ebec, H3C~3A7~Canada}

\begin{abstract}
Coin flipping is a cryptographic primitive in which two 
spatially separated players, who in principle do not trust each other,
wish to establish a common random bit.
If~we limit ourselves to classical communication,
this task requires either assumptions on the computational power
of the players or it requires them to send messages to each
other with sufficient simultaneity to force their complete independence.
Without such assumptions, all classical protocols are so that one
dishonest player has complete control over the outcome.
If~we use quantum communication, on the other hand,
protocols have been introduced that limit the maximal bias
that dishonest players can produce.
However, those protocols would be very difficult to implement
in practice because they are susceptible to realistic losses on the
quantum channel between the players or in their 
quantum memory and measurement apparatus.
In~this paper, we introduce a novel quantum protocol
and we prove that it is completely impervious to loss.
The protocol is \emph{fair}
in the sense that either player has the same
probability of success in cheating attempts at
biasing the outcome of the coin flip. 
We~also give explicit
and optimal cheating strategies for both players.
\end{abstract}
\maketitle

\section{Introduction}
Coin flipping by telephone was first intro\-duced
with these words by Manuel Blum in 1981:
``Alice and Bob [\ldots] have just divorced, live in different cities,
want to decide who gets the car''~\cite{Blum:coinflip}.
They agree that the best thing to do
is to flip a coin, but neither of them
trusts the other and they are unable to agree
on a mutually trusted third party to do the flip for them.
More generally, coin-flipping protocols (also known as ``coin-tossing'')
can be used whenever two players need to pick a random bit even though
it could be to the advantage of one of them (or~perhaps both) to choose,
or at least bias, the outcome of the protocol.

The original coin-flipping protocol introduced by Blum is
\emph{asynchronous} in the sense that it consists of a sequence of rounds
in which the two players alternate in sending messages to each other.
The security of Blum's protocol depends on the assumed difficulty
of factoring large numbers.
Such an assumption is of course of little value in our quantum world,
owing to Peter Shor's algo\-rithm~\cite{shor}, but classical coin flipping can be
based on more general one-way functions, which could potentially be immune to
quantum attacks. Nevertheless, any coin-flipping
protocol that takes place by the asynchronous transmission of classical
messages has the property that one of the \mbox{players} has complete control
over the outcome, given sufficient computing power.
In~the best case, such protocols can be computationally secure,
and even \emph{that} depends on unproven computational complexity
assumptions.

Unconditionally secure classical coin-flipping protocols are possible
in the \emph{synchronous} model, in which the players are requested
to send messages to each other with sufficient simultaneity to force
their complete inde\-pend\-ence. Such protocols are called
\emph{relativistic} because special rela\-tiv\-ity must be invoked to prevent
Alice from \mbox{waiting} to receive Bob's message before choosing
her own (and~vice versa). Rela\-tiv\-istic protocols must be implemented
carefully \mbox{because} their security depends crucially on the physical distance
between the players, and either of them could try to fool the other by pretending
to be \mbox{farther} away than they really are. Such cheating \mbox{attempts} can be
thwarted if each player has a trusted agent near the other player~\cite{kent:relativity}.
For the rest of this paper, we only consider asynchronous protocols
and ``coin-flipping \mbox{protocol}'' will systematically mean
``asynchronous coin-flipping \mbox{protocol}''.

In \emph{quantum} coin-flipping protocols, Alice and Bob are allowed to
exchange quantum states. Such protocols were first investigated
in 1984 by Charles H. Bennett and Gilles Brassard~\cite{bb84}.
In~that paper, a protocol was presented and it was shown that
``ironically [it] can be subverted by use of a still subtler quantum phenomenon,
the Einstein-Podolsky-Rosen paradox'',
making it the first use of entanglement~\cite{EPR35} in quantum cryptography.
We~shall refer to it henceforth as the ``BB84 protocol''
(not to be confused with the better-known quantum key distribution
protocol introduced in the same paper).
The question was left open: Can there be a perfect quantum coin-flipping protocol?
The~proof that this is impossible
was given more than a decade later by Hoi-Kwong Lo and
Hoi Fung Chau~\cite{Lo:nogo},
whose result was further clarified
by \mbox{Dominic} Mayers, Louis Salvail
and Yoshie Chiba-Kohno~\cite{MSC}.
Never\-theless, if quantum coin-flipping protocols cannot be perfect,
can they at least be better than anything classically possible?

To make this question more precise, we say
that one player enjoys an \mbox{$\eps$-bias} if a cheating strategy
exists by which that player could choose either bit
and influence the outcome of the protocol to be that bit
with probability at least~$\frac12+\eps$,
assuming that the other player follows the protocol honestly.
This definition is \emph{unconditional} in the sense that
we allow the would-be cheater to enjoy unlimited computational power
and a technology limited only by the laws of physics.
The~\emph{bias} of a protocol is the largest value of~$\eps $
so that at least one player \mbox{enjoys} an \mbox{$\eps $-bias}.
A~\emph{perfect} protocol would be one whose bias is~$0$,
but they cannot exist, classical or quantum.
At~the other end of the spectrum, a protocol whose bias is~$0.5$
is considered to be \emph{completely broken}.
All~classical protocols are completely broken by this definition,
and so is the BB84 quantum protocol.
The question at the end of the previous paragraph was therefore:
Is~there a quantum coin-flipping protocol whose bias is
strictly less than~$0.5$?
\looseness=+1

The first such protocol was discovered in 2000 by
Dorit Aharonov, Amnon Ta-Shma, Umesh Vazirani and Andrew C.-C. Yao~\cite{ATVY},
who proved that the bias of their protocol (ATVY) is at most \mbox{$\sqrt{2}-1 < 0.42$}
(without any claim concerning the tightness of their bound).
It~was subsequently proven by Robert W. Spekkens and Terry Rudolph~\cite{rud-spek-BB84}
that the ATVY protocol is even better than its inventors had thought:
its bias is in fact exactly \mbox{$\sqrt{2}/4 < 0.36$}.
In~the same paper, Spekkens and Rudolph gave an amazingly simple
coin-flipping protocol that achieves the same bias
as well as another one whose bias is merely \mbox{$(\sqrt{5}-1)/4 < 0.31$}.
According to an earlier paper of theirs~\cite{rud-spek-DCB},
that's the smallest bias possible for a coin-flipping protocol in which the
quantum communication is limited to a single qubit.

Meanwhile, Andris Ambainis~\cite{ambainis} and, independently,
Spekkens and Rudolph~\cite{rud-spek-DCB} discovered quantum coin-flipping
protocols whose bias~$0.25$ is even smaller, but they require the transmission
of a qu\emph{trit} (or~one qubit and two qutrits in the case of Spekkens and Rudolph).
On~the other hand, Alexei Kitaev~\cite{kitaev} proved that no
quantum coin-flipping protocol can have a bias below
\mbox{$(\sqrt{2}-1)/2 \approx 0.21$}.
Very recently, Andr\'e Chailloux and Iordanis Kerenidis~\cite{Cha-Ker}
have announced a quantum coin-flipping protocol whose bias is  arbitrarily close
to Kitaev's bound, but it requires an unlimited
number of rounds of interaction as it \mbox{approaches} this bound.

Despite the theoretical success of quantum coin-flipping protocols, compared
to classical protocols, severe practical problems inherent to their implementation
have been discussed by Jonathan Barrett and Serge Massar~\cite{MassarPRA},
who argued that quantum coin flipping is problematic
in any realistic scenario in which noise and loss
can occur in the processing (preparation, transmission
and measurement) of quantum information. 
For this reason, they proposed random bit-\emph{string} generation \mbox{instead} of
\emph{single-shot} coin flipping. However,
this is not inter\-esting from a quantum
cryptographic perspective \mbox{because} the same goal can be achieved
with purely classical means~\cite{string-flip}.

In~a subsequent paper (NFPM)
written in collaboration with Anh Tuan Nguyen, Julien Frison and Kien Phan Huy,
Massar has defined a \emph{figure of merit} on which quantum coin-flipping
protocols can outperform any possible classical protocol even in a realistic
setting and they have verified their concept experimentally~\cite{NFPM}.
Even though their protocol is not broken in the presence of loss, however,
Alice can choose the outcome with near certainty in a realistic setting.
We~claim that, in order to be of practical use, a protocol should be
\emph{loss tolerant}, which we define as being completely impervious to loss
of quantum information.
In~this sense, the NFPM protocol is not loss tolerant because
its bias increases asymptotically towards 0.5 as losses \mbox{become} more and
more severe, which is unavoidable in practice (with current technology)
over increasing distance between Alice and Bob.

In this paper, we concentrate on this most likely source of imperfection in actual
implementations, namely \emph{losses}.
With the exception of the NFPM protocol mentioned above
(which is not loss tolerant),
all~previously proposed quantum coin-flipping protocols
\mbox{become} \emph{completely} insecure even in the absence of noise
as soon as the quantum channel between Alice and Bob is lossy.
We~intro\-duce the first loss-tolerant quantum coin-flipping protocol.
We~prove that our protocol is \emph{fair} in the sense that either Alice or Bob
can enjoy a bias of exactly $0.4$ with an optimal cheating strategy, independently
of the channel's transmission and other sources of losses,
provided quantum infor\-ma\-tion that is not lost is
prepared, transmitted and measured faithfully.

After this Introduction, the structure of the paper is as follows.
We~begin in Section~\ref{BB84:prot} with a review of the original
1984 quantum coin-flipping protocol of Bennett and Brassard~\cite{bb84}
and we explain why it is completely vulnerable to a so-called EPR-attack.
This is interesting not only for historical reasons, but also because
our novel loss-tolerant protocol follows the same template.
Section~\ref{AmbProt} reviews perhaps the most famous of all quantum coin-flipping protocols,
due to Ambainis~\cite{ambainis}, whose theoretical bias is $0.25$.
However, we demonstrate in Section~\ref{AmbWeak}
that the security of that protocol is completely compromised in the presence of
arbitrarily small channel loss.
Moreover, we argue that this problem is \mbox{inherent} to the protocol
in the sense that it cannot be \mbox{repaired} with small corrections.
(The same would be true of the \mbox{$0.25$-bias} qutrit-based
coin-flipping protocol due to Spekkens and Rudolph~\cite{rud-spek-DCB}
as well as of their optimal single-qubit protocol~\cite{rud-spek-BB84}.)
This is due to the notion of \emph{conclusive measurements},
which we review in Section~\ref{mmt}.
We~show in Section~\ref{ourprot} how to combine the strengths
of the original BB84 protocol with those
of the ATVY protocol (which is not loss tolerant either in its published form)
to finally achieve loss tolerance in quantum coin flipping
and we analyse the security of our protocol.
Conclusions and open problems are presented in Section~\ref{concl}.
\looseness=+1

\section{The BB84 Protocol}\label{BB84:prot}

We review the original BB84 quantum coin-flipping protocol as well
as the way it can be broken~\cite{bb84}.
Here are the so-called BB84 states:
\begin{eqnarray*}
\left.\begin{array}{l}\ket{\psi_{0,0}}=\ket{0}\phantom{\ket{+}}\!\!\!\!\!\!\cr
\phantom{\framebox[0mm]{\large X}}\ket{\psi_{0,1}}=\ket{1}\phantom{\ket{+}}\!\!\!\!\!\!\cr
\end{array}\right\}&& a=0\nonumber\\[1ex]
\left. 
\begin{array}{l}
\ket{\psi_{1,0}}=\ket{+}\phantom{\ket{0}}\!\!\!\!\!\!\cr
\phantom{\framebox[0mm]{\large X}}\ket{\psi_{1,1}}=\ket{-}\phantom{\ket{0}}\!\!\!\!\!\!\cr
\end{array}\right\}&& a=1\;,
\end{eqnarray*}
where \mbox{$\ket{\pm}= (\ket{0} \pm  \ket{1})/\sqrt{2} $}.
We say of $\ket{\psi_{a,x}}$ that $a$ is the \emph{basis} and $x$ is the \emph{bit}.
We~define measurement bases
\begin{equation}\label{eq:basis}
\basis_a = \{ \ket{\psi_{a,0}}, \ket{\psi_{a,1}} \}
\end{equation}
for \mbox{$a \in \{0,1\}$}.
In~the full BB84 quantum coin-flipping protocol~\cite{bb84}, Alice would prepare
and send Bob a large number of qubits, all in the same randomly chosen basis~$a$.
To~emphasize the essential features of the protocol, however, we outline below a
simplified version in which a single qubit is used.
\begin{enumerate}
\item\label{stepone} Alice prepares one of the four BB84 states 
$\ket{\psi_{a,x}}$ with basis $a$ and bit $x$ chosen independently at random;
she transmits that qubit to Bob.

\item\label{stepthree} Bob chooses a random \mbox{$\hat{a} \in \{0,1\}$}
and measures the received qubit in basis $\basis_{\hat{a}}$;
let~Bob's result be~$\hat{x}$.

\item\label{steptwo} Bob sends a randomly chosen bit $\guess$ to Alice.

\item\label{stepfour} Alice reveals her original $a$ and $x$ to Bob.

\item\label{stepfive} If \mbox{$a=\hat{a}$} and \mbox{$x \neq \hat{x}$},
Bob aborts the protocol, \emph{calling Alice a cheater};
if~\mbox{$a\neq\hat{a}$}, Bob has no way to verify Alice's honesty.

\item\label{stepsix} If Bob did not abort the protocol, the outcome of the coin flip
is \mbox{$a \oplus \guess$}, where ``$\oplus$'' denotes the sum modulo~$2$
(also known as the ``exclusive~or'').

\end{enumerate}

In this protocol, Bob cannot cheat at all.  The 
only strategy for a cheating Bob would be
to make an educated guess on Alice's choice of~$a$
before deciding on the~$\guess$ to send her at step~\ref{steptwo},
so as to bias the coin-flip outcome \mbox{$a \oplus \guess$}.
However, $a$ corresponds to Alice's random choice of basis.
Hence, the state $\rho_a$ received by Bob at step~\ref{stepone} is either
\mbox{$\rho_0 = \frac{1}{2} \ketbra{0}{0} + \frac{1}{2}\ketbra{1}{1}$}
or
\mbox{$\rho_1 = \frac{1}{2} \ketbra{+}{+} + \frac{1}{2}\ketbra{-}{-}$}.
It~follows from the fact that \mbox{$\rho_0=\rho_1$} that it is impossible
for Bob to guess the value of $a$ better than at random.

On the other hand,
it is obvious that Alice can bias the protocol if she does not mind the risk of
being called a cheater. The simplest approach is to be honest in the first step.
When she receives $\guess$ from Bob, the probability is $50\%$ that
she is happy with the outcome \mbox{$a \oplus \guess$}, in which case she
proceeds honestly with the protocol. On~the other hand, if she is unhappy
with \mbox{$a \oplus \guess$}, she can lie on~$a$ at step~\ref{stepfour}
and send a random~$x$.
In~that case, her probability of being caught
is~$25\%$ since Bob chose \mbox{$\hat{a} \neq a$} with probability~$50\%$,
in which case he obtained \mbox{$\hat{x} \neq x$} also with probability~$50\%$.
All~counted, this allows her to \mbox{enjoy} a \mbox{$0.375$-bias}.
A~slightly more interesting cheat is for her to send
state \mbox{$(\cos \frac{k\pi}{8})\ket0 + (\sin \frac{k\pi}{8})\ket1$}
for a random \mbox{$k \in \{1,3,5,7\}$} at step~\ref{stepone} and declare
the $a$ that suits her wish (with the appro\-pri\-ately chosen~$x$) at
step~\ref{stepfour}.
This \mbox{allows} her to enjoy a bias of
\mbox{$\frac12 \cos^2  \frac{\pi}{8} = (2+\sqrt{2})/8 > 0.42$},
with a probability \mbox{$\frac12 \sin^2  \frac{\pi}{8} = (2-\sqrt{2})/8 < 8\%$}
of being called a cheater.
It~was to make the probability of unde\-tected cheating exponentially small that the
full BB84 protocol required the transmission and measurement of
a large number of qubits~\cite{bb84}.

A much more remarkable kind of cheating is possible for Alice,
as explained in the same paper that introduced the BB84 protocol itself~\cite{bb84},
which allows her to break the protocol completely
(i.e.~enjoy a \mbox{$0.5$-bias}) with no fear of ever being caught.
Let~us say she wishes the outcome of the coin flip to be bit~$c$.
Instead of sending a legitimate BB84 state
or any other pure state at step~\ref{stepone}, Alice sends half an EPR
pair \mbox{$\ket{\Psi^-} =(\ket{01}-\ket{10})/\sqrt{2}$}
to Bob and keeps the other half for herself. 
She~waits until step~\ref{steptwo}, when she learns Bob's choice of~$\guess$,
to measure in basis \mbox{$a=c \oplus \guess$} the half she had kept;
let~$x$ be her measurement outcome.
This tells her that Bob has obtained (or~will obtain, if he has not yet measured)
$\hat{x}= 1 \oplus x$ in case he has measured (or will measure) in basis~$a$.
(Alice does not care about the value of $\hat{x}$ if Bob chooses to measure
in the other basis.)
Hence, she can always obtain her desired outcome by sending those $a$
and $\hat{x}$ to Bob in step~\ref{stepfour}.
\looseness=+1

As subsequently discovered independently by Mayers \cite{Mayers:nogo}
and by Lo and Chau~\cite{Lo:nogoBC} in the context of quantum bit commitment,
this kind of cheating is \emph{always} possible for Alice in any quantum
coin-flipping protocol that has
the structure of the BB84 protocol, regardless of the actual set of quantum states, whenever
the density matrices used to signal \mbox{$a=0$} or \mbox{$a=1$} at step~\ref{stepone}
are identical (\mbox{$\rho_0=\rho_1$}).
This is due to the striking quantum process known as ``\mbox{remote} \mbox{steering}'',
discovered by Erwin Schr\"odinger~\cite{schrodinger} as early as 1936
and better known as the HJW \mbox{Theorem}~\cite{hjw}.
(See~Ref.~\cite{Kirkpatrick} for an entertaining history of this \mbox{theorem}.)
Note that the remote steering attack works just as well if
Bob postpones his meas\-ure\-ment until after
Alice reveals $a$ and~$x$,
and \emph{almost} just as well if $\rho_0$ and $\rho_1$, although different,
are exponentially indistinguishable~\cite{chris_jeroen}.
This last remark caused the demise of the bit commitment scheme
proposed in Ref.~\cite{BCJL} and we shall henceforth not differentiate
between density matrices that are equal and those that are
merely exponentially indistinguishable.

\section{Ambainis' Protocol}\label{AmbProt}

In order to escape the remote steering attack, Ahar\-onov, Ta-Shma, Vazirani and
Yao~\cite{ATVY} introduced a coin-flipping protocol in which~\mbox{$\rho_0 \neq \rho_1$}.
To~reduce even further Alice's possible bias, they shuffled the order of
steps~\ref{stepthree}, \ref{steptwo} and~\ref{stepfour} so that
Bob delays his measurement of Alice's supplied state until after
she tells him what she claims to have sent. In~this way, he
can measure systematically in the \mbox{declared} basis rather than having to measure
in a randomly chosen basis~$\hat{a}$ whose outcome had probability $50\%$ of
being useless.
This allowed ATVY to design a quantum coin-flipping protocol with bias  \mbox{$\sqrt{2}/4 < 0.36$},
as subsequently proven by Spekkens and Rudolph~\cite{rud-spek-DCB}.

To~achieve his smaller bias of~$0.25$, Ambainis used the following states on qutrits
(rather than on qubits):
\begin{eqnarray*}
\left.\begin{array}{l}\ket{\phi_{0,0}}=\frac{1}{\sqrt{2}} \ket{0}+\frac{1}{\sqrt{2}} \ket{1}\cr
\phantom{\framebox[0mm]{\large X}}\ket{\phi_{0,1}}=\frac{1}{\sqrt{2}}\ket{0}-\frac{1}{\sqrt{2}}\ket{1}\cr
\end{array}\right\}&& a=0\nonumber\\[1ex]
\left. 
\begin{array}{l}
\ket{\phi_{1,0}}=\frac{1}{\sqrt{2}}\ket{0}+\frac{1}{\sqrt{2}}\ket{2}\cr
\phantom{\framebox[0mm]{\large X}}\ket{\phi_{1,1}}=\frac{1}{\sqrt{2}}\ket{0}-\frac{1}{\sqrt{2}}\ket{2}\cr
\end{array}\right\}&& a=1\,.
\end{eqnarray*}
Again we say of $\ket{\phi_{a,x}}$ that $a$ is the basis and $x$ is the bit.
This time, we define measurement bases
\[
\basis_a' = \{ \ket{\phi_{a,0}}, \ket{\phi_{a,1}}, \ket{2-a} \}
\]
for \mbox{$a \in \{0,1\}$}.
Here is Ambainis' protocol.
\begin{enumerate}
\item\label{Astepone} Alice prepares one of the four Ambainis states 
$\ket{\phi_{a,x}}$ with basis $a$ and bit $x$ chosen independently at random;
she transmits that qutrit to Bob, who stores it in his quantum memory.

\item\label{Asteptwo} Bob sends a randomly chosen bit $\guess$ to Alice.

\item\label{Astepthree} Alice reveals her original $a$ and $x$ to Bob.

\item\label{Astepfour} Bob takes Alice's qutrit out of quantum memory and
measures it in basis~$\basis_a'$;
let Bob's result be~$\hat{x}$.

\item If \mbox{$x \neq \hat{x}$}, Bob
aborts the protocol, \emph{calling Alice a cheater}.

\item\label{Astepsix} If Bob did not abort the protocol, the outcome of the coin flip
is \mbox{$a \oplus \guess$}.

\end{enumerate}

Alice cannot use remote steering to gain complete control over the outcome of
the coin flip because the density matrices
that could be \mbox{received}
by Bob at step~\ref{Astepone}, corresponding to her choice $a$ of basis,
\begin{equation}\label{rhoAmb}
\begin{array}{l}
\rho_0 = \textstyle \frac{1}{2} \ketbra{\phi_{0,0}}{\phi_{0,0}} + \frac{1}{2}\ketbra{\phi_{0,1}}{\phi_{0,1}}
=\left(
\begin{array}{ccc}
\pbfrac{1}{2} & 0 & 0 \\
0 & \pbfrac{1}{2} & 0 \\
0 & 0 & 0 \end{array}
\right) \\
\text{and}\\
\rho_1 = \textstyle \frac{1}{2} \ketbra{\phi_{1,0}}{\phi_{1,0}} + \frac{1}{2}\ketbra{\phi_{1,1}}{\phi_{1,1}}
=\left(
\begin{array}{ccc}
\pbfrac{1}{2} & 0 & 0 \\
0 & 0 & 0 \\
0 & 0 & \pbfrac{1}{2}\end{array}
\right) , \!\!\!
\end{array}
\end{equation}
are distinct.
It is easy to see that Alice can enjoy a \mbox{$0.25$-bias}
if she sends state \mbox{$(2\ket{0} \pm \ket{1} \pm \ket{2})/\sqrt{6}$}
at step~\ref{Astepone}
and declares $a$ and $x$ appropriately at step~\ref{Astepthree}.
The proof that this is Alice's optimal cheating strategy
is nontrivial but has been worked out
in detail by Ambainis~\cite{ambainis}.
On~the other hand, the fact that \mbox{$\rho_0 \neq \rho_1$}
makes it possible for Bob to cheat by measuring Alice's qutrit
before step~\ref{Asteptwo} in order to learn information about $a$
and bias his choice of~$\guess$ accordingly. 
The most obvious strategy for Bob is to measure
Alice's qutrit in the computational basis \mbox{$\{0,1,2\}$},
which allows him to enjoy a \mbox{$0.25$-bias} as well.
The fact that this is Bob's optimal cheating strategy
follows directly from Carl W. Helstrom's optimal measurement
theory~\cite{helstrom}, which we review in Section~\ref{mmt}.

For completeness, we mention that the independently-discovered
\mbox{$0.25$-bias} quantum coin-flipping protocol of Spekkens
and Rudolph~\cite{rud-spek-DCB} requires Alice to choose a
random bit~$a$, prepare a one-qubit-and-two-qutrit entangled state
\[ \textstyle \ket{\Gamma_a} = \ket{a} \otimes \left( \frac{1}{\sqrt{2}} \ket{00} + \frac{1}{\sqrt{2}} \ket{a+1,a+1} \right) \]
and send Bob one of the two qutrits in step~\ref{Astepone}.
Step~\ref{Asteptwo} is the same as in Ambainis' protocol.
In~step~\ref{Astepthree}, Alice sends the qubit and the other qutrit to Bob,
which allows him to verify her honesty in step~\ref{Astepfour} by performing a POVM
on the combined Alice-provided state whose outcome is either $\ket{\Gamma_0}$,
$\ket{\Gamma_1}$ or ``Alice cheated''. Again, the result of the coin flip
is \mbox{$a \oplus \guess$} provided Bob did not catch Alice cheating.
The essential (but not sufficient) reason why this protocol gives the same bias as Ambainis'
is that the density matrix $\rho_a$ received by Bob at step~\ref{Astepone} is
exactly the same in both protocols, as given by Eq.~(\ref{rhoAmb}).

\section{A Practical Vulnerability}\label{AmbWeak}

Even though Ambainis' analysis of his protocol is mathematically
impeccable, there is a practical problem that cannot be neglected
if one is ever to implement such protocols in real life:
there will be unavoidable losses in the quantum channel between Alice and Bob.
This is true in particular if photons are used  to carry quantum information
and if the quantum channel is an optical \mbox{fibre}.
Further losses are to be expected in Bob's quantum memory and detection
apparatus.
The situation is even worse for the \mbox{$0.25$-bias} protocol of Spekkens
and Rudolph because it requires both Alice and Bob to have quantum memory
since they must store one qutrit each between steps \ref{Astepone} and~\ref{Astepthree}
(there is no real need for Alice to store also the qubit since it contains only classical information).
Please remember that one of the main appeals of quantum cryptography, from
its very beginnings~\cite{bb84}, has been to offer protocols \emph{that can be implemented
with current technology}, yet remain secure against any potentially future
attack so long as quantum mechanics is not violated.

It~follows that there is a possibility\,%
\footnote{\,Actually, with current technology, it is more than a mere ``possibility'':
this will occur in the overwhelming majority of cases.}, when Bob tries to measure Alice's
qutrit at step~\ref{Astepfour} of Ambainis' protocol (or the one-qubit-two-qutrit
system in the protocol of Spekkens and Rudolph), that he does not register anything
\emph{even though both Alice and Bob have been entirely honest}.
How should Bob react in this case?
He~could hardly call Alice a cheater if the most likely cause for
loss is in his own detectors!
There seem to be only two reasonable responses from Bob, as pointed out
already by Barrett  and Massar~\cite{MassarPRA} concerning a quantum
coin-flipping protocol inspired by the quantum gambling
protocol of Won Young Hwang, Doyeol Ahn and Sung Woo Hwang~\cite{HAH01}:
Bob can
(1)~accept Alice's declared $a$ and $x$ on faith or
(2)~request Alice to restart the coin-flipping protocol from scratch.
But both these ``solutions'' are unacceptable.

If Alice knows that Bob will believe her on faith in case he gets
no detection signal, she can totally bias the coin flip with the most
maddeningly simple cheating strategy: she does \emph{nothing}
at all during step~\ref{Astepone} and lets Bob ``store'' the empty signal.\,%
\footnote{\,It would be technologically very challenging for Bob to
perform a so-called ``non-demolition measurement'', which would in principle
allow him to confirm the presence of a qutrit without disturbing its state.}
After having received Bob's choice of~$\guess$, she is then free to
``reveal'' whichever $a$ would produce her desired
outcome~\mbox{$a \oplus \guess$}. Since Bob's measurement
of the empty signal will yield nothing at step~\ref{Astepfour},
he has no other choice but to believe Alice on faith.

On the other hand, if Alice and Bob have agreed to restart the
coin-flipping protocol in case Bob does not detect anything at
step~\ref{Astepfour}, it is Bob who can totally bias the coin flip
without any need to manipulate quantum states.
Whenever he receives Alice's qutrit, he does nothing at all with~it
and sends his random choice of~$\guess$.
If~Alice reveals $a$ at step~\ref{Astepthree} so that he is happy
with outcome  \mbox{$a \oplus \guess$}, with probability~$50\%$,
Bob pretends to measure Alice's long-lost qutrit and claims to
be satisfied with her honesty.
But~if he is not happy with the outcome, Bob simply tells Alice
that he has not registered anything and requests a new
instance of the protocol. This continues until the outcome is to
Bob's liking. This cheating strategy cannot be detected by Alice
whenever the expected probability $p$ of registering an outcome when
both Alice and Bob are honest is at most~$50\%$, provided
Bob asks to restart the protocol with probability \mbox{$1-2p$} even when
progressing to step~\ref{Astepsix} would have produced his desired outcome.

Unless Bob has the technological ability to make sure he
received a quantum state from Alice at step~\ref{Astepone}
without disturbing~it, 
and because Alice will not accept to restart the protocol
after she has revealed $a$ and $x$, he has
only one line of defence against her ``send~nothing''
cheating strategy: he~must measure Alice's signal
immediately upon reception and be allowed to ask her
to restart the protocol from scratch (with an independent random
choice of state) until he actually registers a measurement outcome.
This means that we must revert to the original BB84 template
in which Bob measures before Alice reveals $a$ and~$x$.

As~explained at the beginning of Section~\ref{AmbProt},
this will make it easier for a cheating Alice to escape detec\-tion 
since Bob's measurement can no longer depend on her claimed state.
However, provided Bob chooses his measurement basis at random,
he will carry out with probability $50\%$ the same measurement
he would have performed in Ambainis' original protocol.
As~we prove in Subsection~\ref{BobHonest},
it follows that Bob's probability of catching Alice's eventual cheating
is reduced, but not by more than a factor of~2,
compared to the original protocol.
Hence, Alice's bias is at most~$0.375$ rather than~$0.25$.
The important point is that this bias remains below~$0.5$.

Nevertheless, this modification in Ambainis' protocol, which is made
necessary by practical considerations (with current technology),
reopens the door for Bob to completely break the revised protocol!
When Alice tells him she transmitted a qutrit,
Bob measures it immediately in the computational basis
\mbox{$\{\ket0,\ket1,\ket2\}$}.
If~he obtains \mbox{either} $\ket1$ or $\ket2$, Bob \emph{knows} Alice's
choice of~$a$. This allows him to choose $\guess$ so that \mbox{$a \oplus \guess$}
suits his desired outcome.
On~the other hand, if Bob either obtains~$\ket0$
or~if he does not register anything, then he tells Alice that the
transmission has been unsuccessful
and he \mbox{requests} \mbox{another} qutrit from her.
In~effect, the protocol will proceed to the step in which
Bob sends his choice of $\guess$ to Alice \emph{only} when
he already knows Alice's \mbox{earlier} choice of~$a$.
This means that Bob enjoys a bias of~$0.5$ and the
protocol is completely broken.
To~camouflage his \mbox{chicanery}
(in~case Alice might wonder why the measurement does not succeed more often), 
Bob can pretend at the outset that his detectors
are half as efficient as they
really are. Alter\-na\-tively, he could surreptitiously replace
the quantum channel that links him to Alice with a sufficiently better one.

This fatal flaw in any practical implementation of Ambainis' protocol
comes from one simple consideration. Even though the mixed states
$\rho_0$ and $\rho_1$
(see~Eq.~\ref{rhoAmb})
used at step~\ref{Astepone}
by Alice to partially commit to either
\mbox{$a=0$} or \mbox{$a=1$},
respec\-tively, are non-orthogonal,
hence they cannot be distinguished with certainty by Bob all the time,
\emph{they can be distin\-guished conclusively with positive probability}.
\mbox{After} \mbox{reviewing} below the notion of
conclusive measure\-ments,
we introduce our new protocol in Section~\ref{ourprot}
and prove that its bias is exactly $0.4$
even when arbitrarily severe losses are taken into account.

\section{Different Types of Measurements}\label{mmt}

Consider two non-orthogonal density matrices $\rho_0$ and $\rho_1$
(they could be pure states).
There are several figures of merit in measure\-ments
that attempt to distinguish them~\cite{chris_jeroen}.
Helstrom has studied the optimal measurement to 
output a guess that minimizes the error probability~\cite{helstrom}.
Assuming both $\rho_0$ and $\rho_1$ were equally likely \emph{a~priori},
Helstrom's measurement outputs the correct guess with probability
\begin{equation}\label{eq:helstrom}
\textstyle \frac12 + \frac12 \, \D (\rho_0, \rho_1) \, ,
\end{equation}
where
\begin{equation}\label{eq:tracedist}
\textstyle \D (\rho_0, \rho_1) = \frac12 \, \Tr | \rho_0 - \rho_1 |
\end{equation}
is the \emph{trace distance} between $\rho_0$ and $\rho_1$,
``\,$\Tr$\,'' denotes the trace and
\mbox{$|A| = \sqrt{A^{\dagger}A}$}.
In~particular, if $A$ is a diagonal real matrix, then \mbox{$|A|_{ij}=|A_{ij}|$}.

When the spans of $\rho_0$ and $\rho_1$ are distinct,
there \mbox{exists} another type of measurement, known as \emph{conclusive} measurement
or \emph{Unambiguous State Discrimination}  (USD)~\cite{Ivanovic,Dieks,Peres}.
These measurements have three possible outcomes, ``$0$'', ``$1$'' and ``?'',
the latter of which being called the \emph{inconclusive outcome}.
Whenever outcome \mbox{$a \in \{0,1\}$} is obtained, it is guaranteed
that the measured state was indeed~$\rho_a$
(assum\-ing of course that is was either $\rho_0$ or~$\rho_1$ and that there
were no experimental errors).
Furthermore, the probability of \mbox{obtaining} a \emph{conclusive} outcome
(not~``?'')\ must be strictly positive for either input state.

A well-known example of conclusive measurement can
distinguish between $\ket{0}$ and $\ket{+}$ with conclusive outcome probability
\mbox{$1-1/\sqrt{2} > 29\%$}. 
More to the point of our paper, the obvious measurement in computational basis
\mbox{$\{\ket0,\ket1,\ket2\}$} distinguishes between Ambainis'
$\rho_0$ and $\rho_1$ (Eq.~\ref{rhoAmb}) with conclusive probability~$50\%$.
In~general, any coin-flipping protocol that follows the template of BB84
is vulnerable to the attack described in Section~\ref{AmbWeak}
when a conclusive measurement exists between the corresponding
$\rho_0$ and~$\rho_1$.

It is tempting to think that this line of attack will not apply when conclusive
measurements do not \mbox{exist}. However, this is not necessarily the case.
Erika \mbox{Andersson}, Stephen M. Barnett, \mbox{Anthony} Chefles, Sarah Croke,
Claire R. Gilson and John Jeffers
have studied ``Maximum confidence quantum measurements'' (MCQM),
which somehow interpolate between Helstrom and conclusive
measurements~\cite{CB,CABGJ}.
Like conclusive measurements, MCQMs have some probability
\mbox{$p<1$} of yielding the ``?'' incon\-clu\-sive outcome.
However, when the outcome is either ``$0$'' or ``$1$'', it is correct
with probability at least \mbox{$q>0$}.
Helstrom's measurement maximizes $q$ subject to \mbox{$p=0$}
whereas conclusive measurements (when they \mbox{exist}) minimize $p$
subject to \mbox{$q=1$}.
There can be a trade-off between those two probabilities
in the sense that it is sometimes possible to achieve an increase
in~$q$ (compared to the accuracy of Helstrom's measurement)
by tolerating a larger~$p$. 

The relevance of MCQMs is that Bob could use them to increase his coin-flip bias
at the cost of increasing his probability of asking Alice to restart the protocol
(pretending he has not registered an outcome). If~MCQMs exist for
$q$ arbitrarily close to~$1$, Bob can
come as near as he wants to choosing the coin-flip outcome,
provided Alice has enough patience (na\"{\i}vety?)\ to
accept restarting the protocol indefinitely until Bob tells her
that he is ready to continue. 

To demonstrate that this is a legitimate worry, consider an
(admittedly contrived) quantum coin-flipping protocol based on
Ambainis' states (see~the beginning of Section~\ref{AmbProt}),
except that we add
\mbox{$\ket{\phi_{0,2}}=\ket{2}$} and \mbox{$\ket{\phi_{1,2}}=\ket{1}$}.
In~step~\ref{Astepone}, Alice chooses basis \mbox{$a \in \{0,1\}$} at random
with uniform probability, but \mbox{$x \in \{0,1,2\}$} is chosen so that
\mbox{$\pr(x=0)=\pr(x=1)=49\%$} whereas \mbox{$\pr(x=2)=2\%$}.
The density matrix received by Bob would be either
\begin{equation}\nonumber
\rho_0'  = \textstyle
\left(
\begin{array}{ccc}
0.49 & 0 & 0 \\
0 & 0.49 & 0 \\
0 & 0 & 0.02 \end{array}
\right)
\text{~or~}
\rho_1' = \textstyle
\left(
\begin{array}{ccc}
0.49 & 0 & 0 \\
0 & 0.02 & 0 \\
0 & 0 & 0.49\end{array}
\right).
\end{equation}
These two mixed states cannot be distinguished conclusively.
Nevertheless, a measurement in the computational basis yields either $\ket{0}$,
which is interpreted as the inconclusive outcome~``?'',
or it yields either $\ket{1}$ or $\ket{2}$, which are interpreted as either $\rho_0'$
or $\rho_1'$, respectively. This MCQC is inconclusive with probability
\mbox{$p=49\%$}. When it is not inconclusive, however,
the verdict is correct with probability \mbox{$q = 0.49/0.51 > 96\%$}.
This is much better than Helstrom's measurement,
which would \mbox{always} give an answer but be correct only with probability~$73.5\%$
since \mbox{$\D(\rho_0',\rho_1')=0.47$}.
Therefore, a quantum coin-flipping protocol that uses these states would allow Bob
to enjoy a maximal theoretical bias of $0.235$ if we did not take losses into account.
However, Bob's bias becomes larger than $0.46$ if Alice agrees to
restart the protocol whenever he claims to have not registered an outcome,
which he would do whenever his measurement outcome is the inconclusive~$\ket{0}$.

\section{A Loss-Tolerant Protocol}\label{ourprot}

\begin{figure}[bt]
  \centering
  \includegraphics*[width=9cm,angle=0]{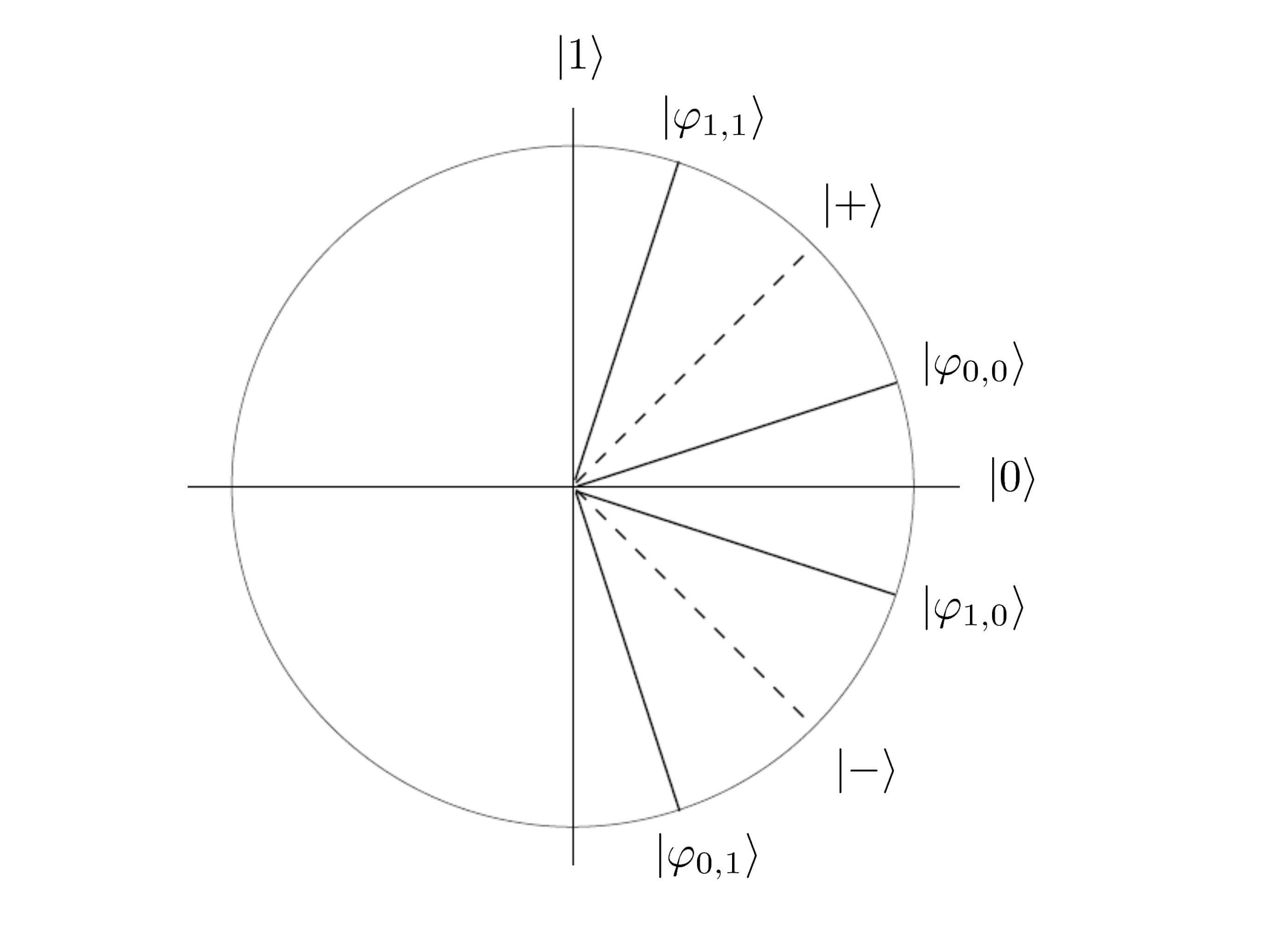}
  \caption{Quantum states used in loss-tolerant protocol.\\
  States such as \mbox{$\ket{\varphi_{0,0}}=\alpha\ket0+\beta\ket1$}
  are represented in the real Cartesian plane by a line between the origin
  and point \mbox{$(\alpha,\beta)$}.}
  \label{fig1}
\end{figure}

Now, we introduce our novel  quantum coin-flipping protocol
and we prove that its bias is exactly~\mbox{$0.4$},
regard\-less of the extent of losses that are unavoidable with current technology.
For this, we use the states introduced by ATVY in their protocol~\cite{ATVY}
but revert to the original BB84 template~\cite{bb84}.
Consider the states
\begin{eqnarray*}
\left.\begin{array}{l}\ket{\varphi_{0,0}}=\alpha \ket{0}+\beta \ket{1}\cr
\phantom{\framebox[0mm]{\large X}}\ket{\varphi_{1,0}}=\alpha \ket{0}-\beta \ket{1}\cr
\end{array}\right\}&& x=0\nonumber\\[1ex]
\left. 
\begin{array}{l}
\ket{\varphi_{0,1}}=\beta \ket{0}-\alpha \ket{1}\cr
\phantom{\framebox[0mm]{\large X}}\ket{\varphi_{1,1}}=\beta \ket{0}+\alpha \ket{1}\cr
\end{array}\right\}&& x=1
\end{eqnarray*}
where $\alpha$ and $\beta$ are real numbers such that
\mbox{$1 > \alpha > \beta > 0$}
and \mbox{$\alpha^2+\beta^2=1$}.
See~Fig.~\ref{fig1}.
As~always, we say of $\ket{\varphi_{a,x}}$ that $a$ is the basis and $x$ is the bit.
We~define measurement bases \emph{mutatis mutandis} as we had done
in Eq.~(\ref{eq:basis}) for the BB84 states:
\begin{equation}\label{double-prime}
\basis_a'' = \{ \ket{\varphi_{a,0}}, \ket{\varphi_{a,1}} \}
\end{equation}
for \mbox{$a \in \{0,1\}$}.
We~shall soon see why we have regrouped the states according to the value of~$x$
rather than that of~$a$ as we had done previously.
Here is our loss-tolerant quantum coin-flipping protocol.

\begin{enumerate}

\item\label{Nstepone} Alice prepares one of the four states 
$\ket{\varphi_{a,x}}$ with basis $a$ and bit $x$ chosen independently at random;
she transmits that qubit to Bob.

\item\label{Nsteptwo} Bob chooses a random \mbox{$\hat{a} \in \{0,1\}$}
and measures the received qubit in basis $\basis_{\hat{a}}''$.
If~his apparatus does not register an outcome,
he requests Alice to start over at step~\ref{Nstepone};
otherwise, let the measurement result be~$\hat{x}$.

\item\label{Nstepthree} Bob sends a randomly chosen bit $\guess$ to Alice.

\item\label{Nstepfour} Alice reveals her original $a$ and $x$ to Bob.

\item\label{Nstepfive} If \mbox{$a=\hat{a}$} and \mbox{$x \neq \hat{x}$},
Bob aborts the protocol, \emph{calling Alice a cheater};
if~\mbox{$a\neq\hat{a}$}, Bob has no way to verify Alice's honesty.

\item\label{Nstepsix} If Bob did not abort the protocol, the outcome of the coin flip
is \mbox{$x \oplus \guess$}.

\end{enumerate}

There are three differences, compared to the original BB84 protocol described in Section~\ref{BB84:prot}:
(1)~in~addition to having been globally rotated to facilitate subsequent analysis,
the states used are more general in the sense that bases $\basis_0''$ and $\basis_1''$
need not be mutually unbiased;
(2)~step~\ref{Nsteptwo} allows Bob to ask Alice to restart the protocol
in case his measurement appa\-ratus fails to register an outcome; and
(3)~the final coin-flip result is \mbox{$x \oplus \guess$} rather than \mbox{$a \oplus \guess$}. 

The first modification will allow us to fine-tune the protocol in Subsection~\ref{fair}.
The second modification makes it possible for the protocol to be loss tolerant
as we have explained in Section~\ref{AmbWeak}. The effect of the third modification,
which corresponds to the main original contribution of the ATVY protocol,
is that two \emph{distinct} density matrices
$\varrho_0$ and $\varrho_1$ (see below) are now used by Alice to partially
commit to either \mbox{$x=0$} or \mbox{$x=1$} by the transmission of
$\ket{\varphi_{a,x}}$ at step~\ref{Nstepone} with a randomly \mbox{chosen}~$a$.
\looseness=+1
\begin{equation}\label{rhoNous}
\begin{array}{l}
\varrho_0 = \textstyle \frac{1}{2} \ketbra{\varphi_{0,0}}{\varphi_{0,0}}
  + \frac{1}{2}\ketbra{\varphi_{1,0}}{\varphi_{1,0}}
= \left(
\begin{array}{cc}
\alpha^2 & 0  \\
0 & \beta^2 \\
\end{array}
\right) \\
\text{and}\\
\varrho_1 = \textstyle \frac{1}{2} \ketbra{\varphi_{0,1}}{\varphi_{0,1}}
  + \frac{1}{2}\ketbra{\varphi_{1,1}}{\varphi_{1,1}}
= \left(
\begin{array}{cc}
\beta^2 & 0  \\
0 & \alpha^2 \\
\end{array}
\right)
\end{array}
\end{equation}

In~sharp contrast, the intuition behind the original BB84 protocol was for Alice to
``commit'' to either \mbox{$a=0$} or \mbox{$a=1$} with a randomly
chosen~$x$, but that was doomed
by Schr\"odinger's remote steering process because the corresponding
density matrices were equal.

Furthermore, we shall see in Subsection~\ref{AliceHonest} that,
contrary to the mixed states $\rho_0$ and $\rho_1$ used in Ambainis' protocol
(see~Eq.~\ref{rhoAmb}),
this time $\varrho_0$ and $\varrho_1$ cannot be distinguished conclusively,
nor even by a maximum confidence quantum measurement better than
Helstrom's measurement, which is the key to loss tolerance.

Although we have presented our protocol as a modification of the original BB84 protocol,
it is useful for analysis purposes to contrast it also with the ATVY protocol.
The key difference between the ATVY protocol and ours is that
they chose to introduce a new template
(used in most subsequent protocols such as Ambainis')
in which Bob stores the quantum state sent by Alice at step~\ref{Nstepone}
in order to
delay measurement until Alice has given him its classical description.
The intention was to make it harder for Alice to cheat
because Bob would know to measure
the state in basis~$\basis_a''$ rather than \mbox{having} to choose a random
basis~$\basis_{\hat{a}}''$. Unfortunately, as we have explained,
it was this feature that made their protocol incapable of tolerating channel
loss unless Bob has the technological ability to detect if a signal has
been received by Alice without otherwise disturbing its state.
\looseness=+1

Next, we prove that the
maximum biases that Alice and Bob can enjoy are
\[ \eps_A = (1+2\alpha\beta)/4
\mbox{~~and~~}
\eps_B= \alpha^2-1/2 \, , \]
respectively, and
we give explicit cheating strategies to achieve those \mbox{biases}.
We~conclude that the choice of $\alpha$ and $\beta$ that
makes those two biases equal corresponds to a fair protocol
whose bias is~0.4.

\subsection{Alice's optimal cheating strategy}\label{BobHonest}

It would be relatively easy to determine Alice's opti\-mal cheating strategy
were she restricted to sending some pure state to Bob in the first step of the protocol.
However, we have learned
from the demise of the original BB84 protocol~\cite{bb84} that it might be to her
dishonest advan\-tage to prepare an entangled state, send
one qubit from it to Bob at step~\ref{Nstepone}, and wait until Bob's
announcement of $\guess$ at step~\ref{Nstepthree}
to measure in the most informative way what she had kept.
This could increase her chances of deciding on her best choice of
$a$ and $x$ to ``reveal'' at step~\ref{Nstepfour} in order to maximize her
probability of successfully biasing the coin flip.
It~is significantly more complicated to take all possible
such strategies into account.
Fortunately, most of the work has already been done by
Spekkens and Rudolph in their thorough analysis
of the ATVY protocol~\cite{rud-spek-BB84}.

Let us begin our analysis by briefly pretending that 
Bob does not measure the state Alice has sent him \mbox{until}
after she reveals her choice of basis and bit, and that he measures
it in that basis.  As we have already pointed out, this becomes
the ATVY protocol.
It~follows directly from the analysis of Spekkens and Rudolph
[choosing what they call $\theta$ in their Eq.~(20) so that
\mbox{$\cos\theta=\alpha$} and \mbox{$\sin\theta=\beta$}]
that any cheating strategy Alice may deploy gives her bias
\begin{equation}\label{SRAlice}
\eps_A' \le \frac{\sin 2\theta}{2} = \sin\theta \cos\theta = \alpha\beta \, .
\end{equation}

Now, to analyse our protocol, we must take into \mbox{account} the fact that we
require Bob to measure Alice's state \emph{before} she tells him
what she claims to have sent.
(This is how we make our protocol  loss tolerant.) 
This difference makes it easier for Alice
to cheat because Bob's measurement cannot be chosen to maximize
his probability of discriminating between her claimed state and
any other state that she might have sent \mbox{instead}.
However, as we show below, this modification does
not reduce Bob's probability of catching Alice 
cheating by more than a factor of~$2$.

Recall that Bob is required by step~\ref{Nstepthree} of the protocol to send
a \emph{random} choice of~$\guess$,
which must be uncorrelated with his former choice $\hat{a}$
of measurement and its outcome~$\hat{x}$.
The crucial observation is that the randomness of~$\guess$
deprives Alice from any information she might otherwise have
obtained from Bob concerning $\hat{a}$ and~$\hat{x}$.
(The~importance of this observation is illustrated in Section~\ref{cunning}.)
It~follows that her choice of which $a$ and $x$ to declare at step~\ref{Nstepfour}
cannot depend on what has happened at Bob's after she transmitted her
quantum state at step~\ref{Nstepone}.
In~particular, \mbox{$\hat{a}=a$} with probability~$50\%$
since $\hat{a}$ is chosen at random by an honest Bob,
in which case he has chosen by chance at step~\ref{Nsteptwo}
precisely the measurement he would have performed in the
ATVY  protocol, had he been allowed 
to wait until Alice's announcement
of $a$ and $x$ before deciding on his measurement.

The above implies that any cheating strategy that Alice might deploy
against our protocol translates into an \emph{identical} cheating strategy
against the ATVY protocol, possibly with 
a different success probability,
since it makes no measurable difference
to Alice whether Bob measures before (as in our protocol) or after
(as in the ATVY protocol) she has to declare $a$ and~$x$.
Now, consider an arbitrary cheating strategy on the part of Alice and
let~$\eps_A$ (resp.~$\eps_A'$) be the bias that she enjoys with this strategy against
our protocol (resp.~the ATVY protocol).

Consider an arbitrary run of our protocol, when Alice uses this cheating strategy.
With probability $50\%$, inde\-pend\-ently from anything else, Bob randomly chooses the same
measurement he would have performed in the ATVY protocol, in which case Alice
succeeds with probability \mbox{$\frac12+\eps_A'$}.
In~the other $50\%$ of the cases, Bob cannot verify the state Alice claims
to have sent and the probability that Alice succeeds is
at most~$1$. All~counted, the success probability of Alice in our protocol is
\[ {\textstyle \frac12}+\eps_A \le {\textstyle \frac12} \left( {\textstyle \frac12}+\eps_A' \right)+{\textstyle \frac12} \times 1 
 = {\textstyle \frac34} + {\textstyle \frac12} \eps_A'  \, . \]
It~follows that
\begin{equation}\label{alicebias}
 \eps_A \le \frac {1+2 \eps_A'}{4} \le \frac {1+2 \alpha\beta}{4} \, ,
\end{equation}
where the last inequality follows from Eq.~(\ref{SRAlice}).

We now proceed to show that this bound can be saturated.
Surprisingly (when we recall the demise of the BB84 quantum coin-flipping protocol),
the optimal cheating strategy for Alice does \emph{not} require her to
prepare an entangled state from which she would send one qubit at step~\ref{Nstepone}.
Instead, it suffices for her to send either
\mbox{$\ket{+} = (\ket{0} + \ket{1})/\sqrt{2}$} or \mbox{$\ket{-} = (\ket{0} - \ket{1})/\sqrt{2}$}
to Bob. Let us say she sent $\ket{+}$ (the~other case is similar) and \mbox{received}
bit~$\guess$ from Bob at step~\ref{Nstepthree}. If~her desired outcome is~$c$,
she sets \mbox{$x = c \oplus \guess$} and
claims at step~\ref{Nstepfour} to have sent $\ket{\varphi_{x,x}}$ at step~\ref{Nstepone}.
With probability $50\%$, Bob had already chosen \mbox{$\hat{a} \neq x$}, in which case
he cannot catch Alice cheating. With complementary probability  $50\%$, he had chosen
\mbox{$\hat{a} = x$}, in which case Alice escapes detection with probability
\[ |\braket{+}{\varphi_{x,x}}|^2 = \textstyle \left( \frac{1}{\sqrt{2}} \alpha + \frac{1}{\sqrt{2}} \beta \right) ^2
= \displaystyle \frac{(\alpha+\beta)^2}{2}
= \textstyle \frac{1}{2}+\alpha\beta \]
(the last equality is because \mbox{$\alpha^2+\beta^2=1$}).
Putting it all together, Alice obtains her desired outcome with probability
\[  \frac{1}{2} + \frac{1}{2} \left(  \frac{1}{2}+\alpha\beta \right)
= \frac{3+2\alpha\beta}{4} \]
and her bias is therefore
\[  \eps_A = \frac{3+2\alpha\beta}{4} - \frac{1}{2} = \frac{1+2\alpha\beta}{4} \, , \]
which saturates the bound given in Eq.~(\ref{alicebias}).

\subsection{Bob's optimal cheating strategy}\label{AliceHonest}

In the analysis of Bob's bias in quantum coin-flipping protocols, it is usual to
apply Eqs.~(\ref{eq:helstrom}) and~(\ref{eq:tracedist}) to compute the trace distance
$\D(\varrho_0,\varrho_1) = \alpha^2-\beta^2 = 2\alpha^2-1$
in \mbox{order} to determine that the optimal
Helstrom measurement that Bob could perform to best guess Alice's committed bit
(here,~$x$) gives him the correct answer with probability~$\alpha^2$.
From this, we would ``normally'' conclude that Bob's maximal bias is
\mbox{$\alpha^2-1/2$}.

As~we have seen, however, this approach is not appro\-priate in our context
because it does not take into \mbox{account} the eventual possibility for Bob to increase his
bias by making conclusive or maximum confidence quantum measurements
on the state sent by Alice at step~\ref{Nstepone}, so that he could ask her
to restart the protocol whenever he is not sufficiently satisfied with the probability
that his guess of $x$ be correct.

Fortunately, the analysis of Bob's optimal cheating strategy is straightforward in this case.
For~either value of \mbox{$x \in \{0,1\}$} that Alice may have chosen in step~\ref{Nstepone},
Eq.~(\ref{rhoNous}) shows the mixed state $\varrho_x$
that Bob would receive from her. The key observation is that, algebraically speaking,
\mbox{$\varrho_0 = \alpha^2 \ketbra{0}{0} + \beta^2 \ketbra{1}{1}$} and
\mbox{$\varrho_1 = \beta^2 \ketbra{0}{0} + \alpha^2 \ketbra{1}{1}$}.
It~follows that, from a cheating Bob's
perspective, whose only concern is in guessing~$x$ as best as possible
(including the possibility of asking Alice to restart the protocol if
he is not satisfied with his own confidence in his guess),
this is \emph{strictly identical} to an alternative scenario in which Alice
would have sent $\ket{0}$ with probability $\alpha^2$ or $\ket{1}$ with probability $\beta^2$
in case \mbox{$x=0$}, and vice versa in case~\mbox{$x=1$}.
In~other words, this is exactly as if Alice had communicated her choice of~$x$ to Bob
as the \emph{purely classical} signal $\ket{x}$
through a binary symmetric channel with error probability~$\beta^2$.

Seen this way, there is \emph{nothing quantum} about the situation.
Hence, a complete measurement in the computational basis
of the state received by Alice provides Bob with an optimal cheating strategy
since it yields \emph{all} the infor\-mation classically available
in the ``quantum'' signal!
Not~surprisingly, this is indeed precisely Helstrom's measurement. 
In~particular, the outcome of this measurement does not give Bob any indication
on whether or not it would be to his advantage to ask Alice to restart the protocol.
(Note that it is not simply because Alice's signal can be thought of as classical
that there cannot be a conclusive or a maximum confidence measurement;
in~particular, \emph{erasure channels} provide the classical equivalent to
quantum conclusive measurements---but binary symmetric channels don't.)
\looseness=+1

To~summarize, Bob's optimal cheating strategy is to measure Alice's qubit
in the computational basis in \mbox{order} to learn the value of~$x$ with error probability~$\beta^2$.
This \mbox{allows} him to choose $\guess$ appropriately and obtain his \mbox{desired} outcome
\mbox{$x \oplus \guess$}
with complementary probability~$\alpha^2$, thus enjoying his optimal bias
\mbox{$\eps_B = \alpha^2 - 1/2$}.

\subsection{A fair protocol}\label{fair}
In this subsection, we find the 
value of $\alpha$ so that the bias
either player can achieve by cheating is the same.  
We simply need to fulfil the condition
\[
\eps_A=\eps_B \, ,
\]
which for our protocol amounts to
\[
\frac{1+2\alpha\beta}{4} =  \alpha^2-\frac{1}{2} \, ,
\]
subject to \mbox{$\alpha^2+\beta^2=1$}.
Solving this system yields
\[ \alpha = \sqrt{0.9} \mbox{~~and~~} \beta = \sqrt{0.1}  \, , \]
as illustrated in Fig.~\ref{fig1}.
These values correspond to
\[
\eps_A=\eps_B=0.4 \, ,
\]
which defines a fair loss-tolerant quantum coin-flipping protocol whose bias is~$0.4$.
In~other words, either player can obtain a desired outcome with probability $90\%$
by using an optimal cheating strategy, provided of course the other player is honest.
But~what if they both try to cheat?

\subsection{The cunning game}
\label{cunning}

An interesting phenomenon is illustrated if we take the unusual step of
considering the case when both Alice and Bob cheat.
The most \mbox{obvious} example of double-cheating occurs when Alice is convinced
that Bob is so greedy that he will try his best to control the coin flip.
In~this case, Alice can send an honest quantum state 
in step~\ref{Nstepone}.
To~maximize his chances of guessing Alice's choice of~$x$,
Bob measures the state in the computational basis, 
as we have seen.
This allows him to send Alice a value of~$\guess$ that produces
his desired outcome with probability $90\%$
(assuming they use the fair version of the protocol). 
But at the same time, he has lost his ability to verify the honesty of Alice
since he did not measure her qubit in a legitimate basis.
Suspecting this, Alice is free to claim whatever suits her best at step~\ref{Nstepfour},
probably lying about~$x$, and Bob cannot look her in the face and call her a cheater!

For an amusing variation on this theme, consider what happens if both parties
attempt to perform their optimal cheating strategies simultaneously.
In~this case, Alice sends either $\ket{+}$ or~$\ket{-}$, which Bob measures
in basis \mbox{$\{\ket{0},\ket{1}\}$}.
When Bob attempts influencing the coin flip by choosing $b$
as a function of his measurement outcome,
little does he suspect that his choice is in fact totally random:
ironically, Bob follows the honest protocol at step~\ref{Nstepthree}!
\looseness=-1

For a more subtle example,
consider a scenario according to which Bob is Alice's young son.
They agree to flip a quantum coin to deter\-mine who will decide
on the film to be seen tonight: Alice will have the choice if the outcome
is~$1$ and Bob if it is~$0$.
Alice knows that her little rascal will do his best to cheat and win,
but also that he will follow step~\ref{Nsteptwo} properly
(rather than measuring in the computational basis) because he will
not want to relinquish his possibility to catch his mother cheating
at verification step~\ref{Nstepfive}---or~so he thinks.
Hence, Bob's cheating will consist in sending
\mbox{$\guess=\hat{x}$} at step~\ref{Nstepthree} instead of choosing $\guess$
at random. After a calculation, we find that this gives him probability
$82\%$ that \mbox{$\guess=x$},
thus producing his winning outcome \mbox{$x \oplus \guess = 0$}. 

Unbeknownst to Bob, however, his mother wants him to win,
but she does not wish him to know this for fear of undermining
her authority! Assuming she knows her son as well as she thinks,
she proceeds as follows.
At~step~\ref{Nstepone}, she honestly sends some random
state~$\ket{\varphi_{a,x}}$.
When she receives Bob's choice of $\guess$ at step~\ref{Nstepthree},
there are two possibilities. If~\mbox{$\guess=x$}, she can relax and
continue honestly since Bob wins, according to her wish.
On~the other hand, if \mbox{$\guess \neq x$}, then Alice can deduce
that Bob used the wrong basis in his measurement:
\mbox{$\hat{a} \neq  a$}. 
This allows her to ``reveal'' $a$ together with any $\tilde{x}$ of her choice
at step~\ref{Nstepfour} since she knows (or~suspects) that Bob will be
unable to call her a cheater. According to our story, she chooses
\mbox{$\tilde{x}=\guess$} to make Bob happy.

Admittedly, the above tale is unlikely at best.
Nevertheless, it serves the purpose of demonstrating the importance in our
proof of security against Alice (Subsection~\ref{BobHonest}) that Bob's bit $\guess$
be chosen randomly despite the
fact that he has significant information about Alice's $x$ after an honest
measurement at step~\ref{Nsteptwo}. Indeed, our proof relied in a crucial way
on the fact that Alice would have no information on the measurement basis
used by Bob. Our~tale shows how wrong and damaging it would have
been had the protocol asked an \emph{honest} Bob to send a $\guess$ correlated to
his measurement outcome.

\subsection{Side channels}

We have seen that Bob could exploit the loss of quantum information to cheat in previous
protocols, whereas our protocol is loss tolerant.
Nevertheless, our protocol could become susceptible to loss if it were implemented
with insufficient care.
A~\emph{side channel} is any source of information that Bob could exploit about
the quantum state sent by Alice above and beyond its theoretical definition as
a pure qubit.
For~example, the protocol would be obviously insecure if the apparatus used by Alice
to generate her quantum states produced each of the four legit\-imate states as
photons of significantly different wavelengths or spatial position.
These issues have been \mbox{studied} exten\-sively in the context of quantum key distribution.
See~Ref.~\cite{hacking} for a compelling example of successful hacking of a
commercially available apparatus.
\looseness=-1

An interesting example of side channel would \mbox{occur} in a careless implementation
of our quantum coin-flipping protocol if Alice used an attenuated laser pulse
to generate her states, as is done in most current implementations of quantum key distribution.
The problem stems from the fact that  one can distinguish conclusively
between $\varrho_0$ and $\varrho_1$ (Eq.~\ref{rhoNous}) when
a pulse consists of two (or more) identical ATVY states.
For~this, it suffices for Bob to measure one photon in basis $\basis_0''$
and the other in basis~$\basis_1''$ (Eq.~\ref{double-prime}).
If~the two measurements produce the
same outcome, this is necessarily the correct bit~$x$ encoded by Alice
since one of the measurement has been performed in the correct basis.
This occurs with probability \mbox{$(\alpha^2-\beta^2)^2$}.
With the fair states, Bob's probability of conclusive outcome is therefore
a substantial $64\%$ each time a pulse contains two photons,
which means that this \emph{implementation} of
the protocol is completely insecure because Bob can \mbox{request} another
state from Alice until this event \mbox{occurs}.

To make the case of attenuated laser pulses even worse,
a perfectly natural measurement apparatus that an \mbox{honest} Bob could use
in the implementation of our protocol~\cite{BBBGST} would produce such conclusive
outcome with half the probability derived above, namely $32\%$
of the double-photon pulses.
It~follows that Bob, who started
the protocol with pure intentions, might find it difficult to resist sliding
to the dark side when his (\mbox{honest!})\ measurement apparatus reveals
with certainty the value of~$x$ chosen by~Alice.
A~practical solution to this problem is for Alice to use a source of entangled photons,
as discussed in Ref.~\cite{BBBGST}.

\section{Conclusion}\label{concl}

Quantum coin flipping is a cryptographic primitive that has been studied extensively.
Several approaches have been considered since the very beginnings of quantum cryptography.
However, previous protocols are either totally insecure~\cite{bb84},
or they are highly sensitive to (if~not completely broken~by)
technologically unavoidably losses of quantum information on the channel
between the players or in their storage and detection
apparatus~\cite{ATVY,ambainis,rud-spek-DCB,rud-spek-BB84,NFPM,Cha-Ker}.
In~this paper, we introduced the first loss-tolerant quantum coin-flipping protocol,
which means that it is completely impervious to such losses.
We~proved that our protocol is \emph{fair} in the sense that both Alice and Bob
have an optimal cheating strategy \mbox{capable} of producing their desired outcome with
$90\%$ probability of success (assuming the other player is honest)
and we provided those strategies explicitly.

Even though our protocol can tolerate arbitrary loss of quantum information,
it would fail in case of \emph{noise} \mbox{because} it would be impossible
for Bob to know, in case of a mismatch between his measurement outcome and
Alice's claimed state, if that is due to a genuine error
(on~his part or Alice's) or~to Alice's dishonesty.
\mbox{Recall} that this may be unavoidable; indeed \mbox{Barrett} and
Massar have argued that single-shot quantum coin-flipping protocols are
problematic when both noise and loss can \mbox{occur}~\cite{MassarPRA}.

If secure single-shot quantum coin-flipping protocols are indeed impossible
in the presence of noise, is there something useful that quantum coin flipping
can do above and beyond anything classically possible?
We~already know that random bit-string generation~\cite{MassarPRA}
is not the answer given that it can be done classically~\cite{string-flip}.
We~are currently investigating this issue~\cite{BBBGST}
along a different line from that proposed in Ref.~\cite{NFPM}.

Another open question concerns the optimality of our protocol.
Could there be a loss-tolerant quantum coin-flipping protocol
whose bias is smaller than~$0.4$?
Alter\-na\-tively, can we prove that $0.4$ is the smallest bias possible
among all loss-tolerant quantum coin-flipping protocols?
Finally, we mention that this paper was entirely concerned with
the task known as \emph{strong} coin flipping.
There exists a similar task, \emph{weak} coin flipping,
in which only one outcome is favourable for Alice, the other
outcome being the only one favourable for Bob.
Protocols with arbitrarily small bias are known to exist
for weak coin flipping~\cite{mochon}, in stark contrast with Kitaev's lower bound
for the strong case~\cite{kitaev}.
Can \emph{loss-tolerant} quantum weak coin-flipping protocols exist
with arbitrarily small bias?
This question will be addressed in a subsequent paper.

We~have recently implemented our own loss-tolerant protocol
and we have successfully
tested it against Alice's and Bob's opti\-mal cheating strategies.
We~report on those experiments elsewhere~\cite{BBBGST}.
This was in a sense going full circle because it was our wish
to implement Ambainis' quantum coin-flipping protocol that lead
to the realization that we could never succeed and thus to the
development of our new protocol.

\section{Acknowledgements}

The authors are grateful to Hass\`ene Bada for pointing out to us
the fact that Spekkens and Rudolph
had worked out a tight analysis of the ATVY protocol
(which was lacking in the ATVY paper) and to
Serge Massar, Carlos Mochon and Robert Spekkens
for discussing their papers very patiently with~us.
Somshubhro Bandyopadhyay, S\'ebastien Gambs and Tal Mor have helped the authors
concerning the issue of maximum confidence quantum measurements.

G.\,Br.~is supported in part by Canada's Natural
Sciences and Engineering Research Council (\textsc{Nserc}),
the Canada Research Chair program,
the Canadian Institute for Advanced Research (\textsc{Cifar}),
Quantum\emph{Works}
and the Institut transdisciplinaire d'informatique quantique (\textsc{Intriq}).
F.\,B.~is supported in part by the Fonds qu\'eb\'ecois de la recherche
sur la nature et les technologies (\textsc{Fqrnt}),
the Canadian Institute for Photonics Innovations (\textsc{Cipi})
and an \textsc{Nserc} Canada \mbox{Graduate} \mbox{Scholarship}.
N.\,G.~is supported in part by
the Centre d'optique, photonique et lasers (\textsc{Copl}),
Quantum\emph{Works}, \textsc{Nserc},
\textsc{Cipi} and \textsc{Intriq}.

\end{document}